\documentclass[runningheads]{llncs}

\usepackage[T1]{fontenc}

\usepackage{cite}
\usepackage{graphicx}
\graphicspath{{img/}}
\usepackage[misc]{ifsym}
\usepackage[colorlinks=true,allcolors=blue]{hyperref}
\usepackage[dvipsnames]{xcolor}
\usepackage[caption=false]{subfig}

\colorlet{designPatternColor}{blue!66!black}
\makeatletter
\newcommand{\designpattern}[2]{\noindent\protected@write \@auxout {}{\string \newlabel {#1}{{\color{designPatternColor}#2}{\thepage}{#2}{#1}{}} }\hypertarget{#1}{{\color{designPatternColor}#2}}}
\makeatother

\newcommand{\authorMetadata}{Conner Bradley, David Barrera}
\newcommand{\metaTitle}{Towards Characterizing IoT Software Update Practices}

\hypersetup{pdfinfo={
Title={\metaTitle},
Author={\authorMetadata}
}}

\begin{document}
\title{\metaTitle}
\author{Conner Bradley\textsuperscript{(\Letter)} \and David Barrera}
    \authorrunning{C.\ Bradley et al.}
    \institute{School of Computer Science, Carleton University, Ottawa, Canada
    \email{connerbradley@scs.carleton.ca}}

    \maketitle

\begin{abstract}
    Software updates are critical for ensuring systems remain free of bugs and vulnerabilities while they are in service. While many Internet of Things (IoT) devices are capable of outlasting desktops and mobile phones, their software update practices are not yet well understood, despite a large body of research aiming to create new methodologies for keeping IoT devices up to date. This paper discusses efforts towards characterizing the IoT software update landscape through network-level analysis of IoT device traffic. Our results suggest that vendors do not currently follow security best practices, and that software update standards, while available, are not being deployed. 
\end{abstract}     

\section{Introduction}~\label{sec:intro}
Consumer Internet of Things (IoT) devices have gained significant popularity in recent years, resulting in a revolution of IoT devices used in many applications. IoT devices are typically resource-constrained and require specialized operating systems and software stacks depending on their application~\cite{bellman2019analysis}. Due to the unique resource constraints of IoT devices, device vendors have to either design their software update infrastructure and supporting applications from scratch or use an integrated third-party solution\footnote{Such as Microsoft Azure IoT, or Amazon Web Services IoT.} which has historically shown to be inconsistent and vulnerable~\cite{cloudIoTSecuritySurvey}. Software update systems are well understood and widely available on general-purpose computers and servers~\cite{bellissimo_secure_nodate}; however, there is very little insight and research into how these vendor-specific IoT software update systems work due to a lack of standardization in the IoT space~\cite{zandberg_secure_2019,bettayeb_firmware_2019}. Our goal is to characterize how typical consumer IoT devices query for and retrieve software updates, and evaluate the security of these techniques as used by prominent IoT vendors. 

A unique challenge for deployed IoT devices is their expected lifespan. Typical personal computers have a relatively short lifespan compared to an IoT device, which is expected to behave in an appliance-like fashion with minimal (if any) downtime. Personal computers may get replaced in 5-10 years if the hardware cannot keep up with current software demands. In contrast, an IoT device such as a smart thermostat may be expected to run for decades before being replaced. With the constant evolution of technology, device vendors have the additional challenge of providing a secure implementation of their software on potentially outdated hardware.

We hypothesize that suboptimal update intervals from IoT device vendors may further weaken IoT update systems. For example, device libraries such as the crucial OpenSSL library were analyzed during a study of 122 IoT device firmware files, which revealed several vendors failed to patch OpenSSL in their IoT devices after critical vulnerabilities were released~\cite{zhang_capture_2021}. Device vendors took months to supply an updated system image with a patched OpenSSL version, and one vendor took nearly 1,500 days to patch the critical vulnerability. Failing to update critical libraries causes these devices to gain a larger attack surface that could potentially be leveraged by bad actors to trick the device into downloading malware~\cite{wang_inside_2017} or to bypass security measures that are in place to prevent the device from loading modified firmware~\cite{cui_when_nodate,bettayeb_firmware_2019}.

In recent years there have been many proposals for secure software update systems that are designed for IoT~\cite{boudguiga_towards_2017,he_securing_2019,zhang_capture_2021} and related cyber-physical systems~\cite{karthik_uptane_nodate, nikitin_chainiac_nodate}; however, there is no research (to our knowledge) aiming to broadly understand the IoT software/firmware update landscape in consumer IoT devices. 

Our primary focus is identifying software updates being requested and taking place. The benefits of this can be leveraged in various contexts:
Network-level update detection can be used as independent feedback to end users that their devices are being updated regularly -- an IoT device vendor may promise to publish security patches for their IoT devices, but not deliver on that promise~\cite{zhang_capture_2021}.
In an enterprise context, administrators may want to apply the principle of least privilege to fleets of IoT devices. Certain IoT devices do not need continuous access to the open internet as most devices can function exclusively with LAN connectivity to a central hub or other devices. The only edge case to this is checking for updates and downloading them. If an active firewall can detect update-related traffic from IoT devices, it can adjust rules to (1) allow the IoT device to download an update from the internet, and (2) log the update instance.

The research contributions in this paper are:
\begin{itemize}

    \item The first in-depth analysis of consumer IoT network traffic to identify software update communications. We identified design patterns used in several IoT devices and found vulnerabilities that could be exploited.

    \item A case study of software update schemes and practices that we identified through our methodology. Devices featured in our case study distribute software updates over HTTP with no tamper-resistant protection mechanisms added on. One of the devices identified in the case study provides a happy medium between update transparency and security.
    
    \item An event-based characterization of when IoT devices update. We contextualize the various conditions that lead to an IoT device performing updates. For example, power cycling an IoT on is highly likely to trigger an update check.

\end{itemize}

\section{Methodology}
Our research objective is to understand and characterize how and when IoT devices perform software updates. To accomplish this, we build a network traffic analysis system that identifies and analyzes software update requests and responses from IoT devices. We aim for the system to be vendor-agnostic, requiring no \textit{a priori} knowledge about the IoT vendor's infrastructure or devices. The system should also identify updates across multiple independent cloud vendors, which are relied upon heavily in IoT. 

To accomplish this, we analyze network traffic from a 2019 Internet Measurements Conference (IMC) paper by Ren et al.~\cite{ren_information_2019} which actively captured traffic from 81 IoT devices. These 81 devices were located in two geographic regions; 46 in the US, 35 in the UK, and 26 common devices across both regions. In total, the dataset contains packet captures from 55 unique devices. Collected data was harvested at network gateways, but no form of middle-person attack was done on TLS traffic which precludes peering into an encrypted device communication. Therefore, in this paper, we rely exclusively on extractable HTTP traffic for identifying software updates. Additionally, we harvest metadata from the TLS handshakes to gain insight into the security of the secure communication channels used by these devices. 

\subsection{Data Extraction}\label{sec:DataExtraction}

In total, the dataset of packet captures from Ren et al.\ is 13 GB in size, which includes $37,744$ packet captures recorded by the automated test system and $611$ unsupervised experiment packet captures, yielding a total of $38,355$ packet captures. We do not separate traffic by geographic region as Ren et al.\ found very negligible differences in region-specific traffic~\cite{ren_information_2019}. 

To identify network traffic related to software updates, we hypothesize that update interactions between an IoT device and vendor cloud follow a structured schema. If the schema is human-readable (e.g., JSON, XML, etc.) there will be keywords contained inside indicating some update-related information, such as a firmware version. We initially searched for a single keyword ``update'', which led us to build a corpus of update-related keywords: update, upgrade, firmware, software, and download.

\begin{figure}
    \center
    \includegraphics[width=0.90\textwidth]{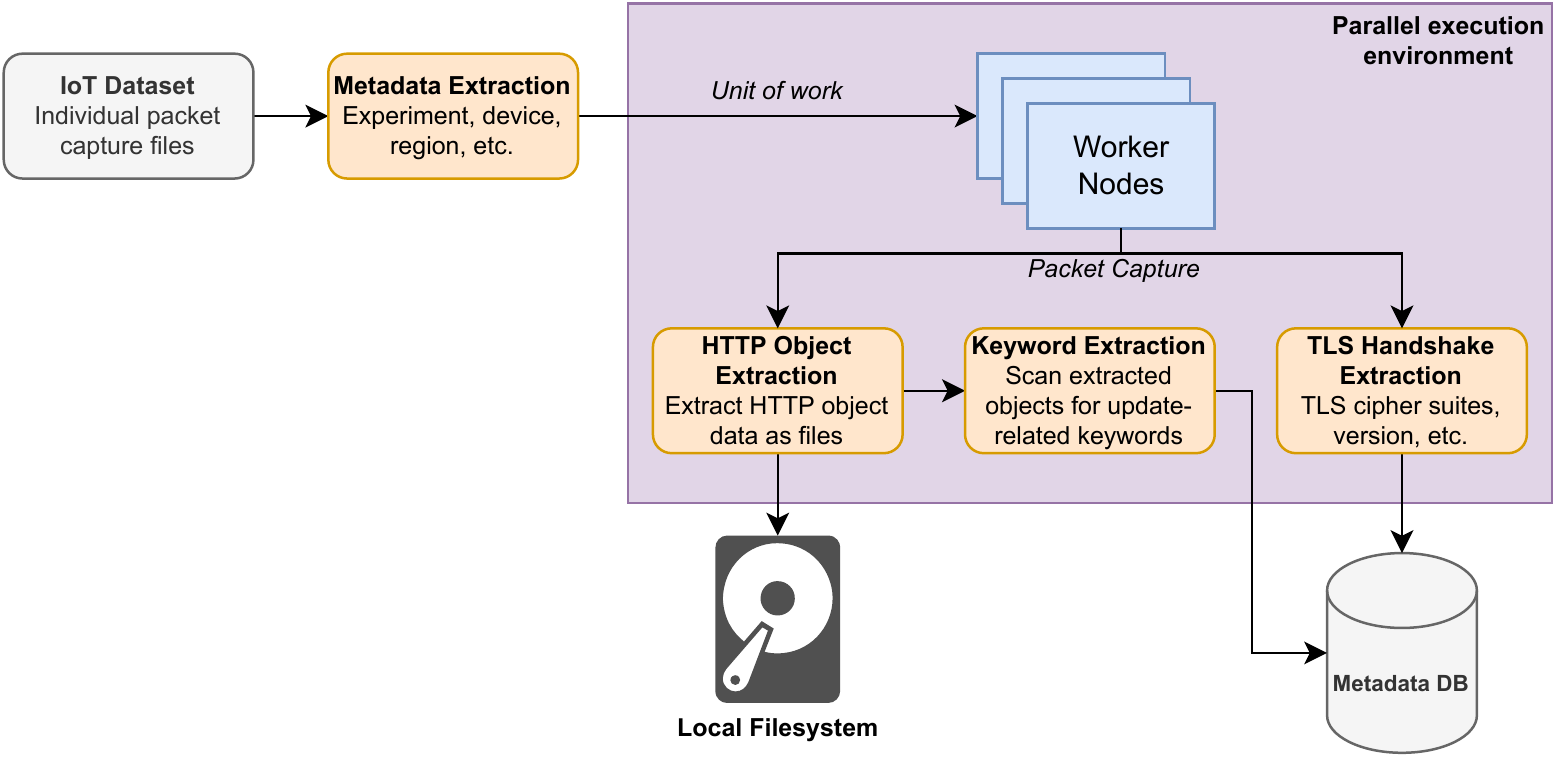}
    \caption{Our data extraction pipeline: starting with the IoT packet capture dataset, we extract metadata for each packet capture to represent a given packet capture as a unit of work. We process each packet capture in parallel, extracting HTTP objects to the local filesystem, TLS Handshakes, and update-related metadata. All extracted metadata is stored relationally in a metadata database for further analysis, and HTTP objects are stored on the local filesystem.}
    \label{ExtractionPipeline}
\end{figure}

These keywords will be the basis we use for identifying update-related traffic; however, manually searching through files will not scale to the number of devices we have. Therefore, we developed a parallel network traffic processing pipeline (see Figure~\ref{ExtractionPipeline}) that manages network traffic metadata and HTTP object extraction. The pipeline design is compatible with distributed data processing frameworks such as Apache Spark, and works on the dataset as follows:

\noindent\textbf{Metadata Extraction}: We extract metadata representing the packet capture. This includes the specific sub-dataset, region, experiment type (e.g., power on, interact with the device, etc), and device name. The extracted metadata is saved to a metadata database and used for later steps in the pipeline.

\noindent\textbf{Parallelization}: We parallelize the extraction of metadata and HTTP objects on a per-packet capture basis. The parallelization is done by assigning each packet capture to a worker node, and the worker node performs the following steps on each packet capture individually. In practice a parallelization approach is not needed; however, passive analysis of a large amount of packet captures warrants the speedup gains of parallelization.

\noindent\textbf{HTTP Object Extraction}: We extract all HTTP payloads from a given packet capture. The HTTP payload data is of particular interest as it provides us insight into any files transferred along with any web service interactions.

\noindent\textbf{TLS Handshake Extraction}: We then extract TLS client and server hello data using a modified version of \texttt{pyshark}\footnote{\href{https://github.com/KimiNewt/pyshark}{https://github.com/KimiNewt/pyshark}}. Our modified version of \texttt{pyshark} supports extracting an extended set of TLS handshake metadata, including the ciphers advertised in the TLS client hello and server hello handshake. In total, we return a list containing every TLS handshake, including the TLS version, TLS handshake type, and a list of cipher suites. The TLS cipher suite data is used to determine if devices are adequately securing communication channels against TLS-related attacks.

\noindent\textbf{Keyword Extraction}: For each of the extracted HTTP objects, we scan for the aforementioned update-related keywords by performing a case-insensitive search for all of the keywords. A keyword occurrence flags a packet capture related to a software update. Counts of keyword occurrences are saved to the metadata DB for future analysis.

The data-extraction pipeline operates per packet capture in parallel. On a test VM with 24 virtual processors, 64 GB of RAM, and a solid-state drive, we were able to run the extraction pipeline on $38,355$ packet captures in over 60 minutes, with approximately 10 packet captures processed per second. Without a parallel approach, our extraction pipeline would have taken over 24 hours to complete.

\subsection{Data Analysis}

Using the metadata that corresponds to the packet capture, we can perform extended analysis on the packet capture that had been flagged as having update-related traffic. After identification of these packet captures, we inspect the HTTP response data to look for any update endpoints or update artifacts. Ideally, we should find no update-related artifacts in HTTP responses, as this would imply these files are transmitted over an insecure channel.

Device vendors \textit{should} be protecting their firmware from being tampered with regardless of the transfer protocol being used: if a vendor uses only TLS to secure their updates in transit, the compromise of a single cryptographic key is the only requirement to jeopardize the integrity of the vendor's update system~\cite{samuel_survivable_2010}.

Analyzing IoT update interactions by raw traffic can be misleading as it does not consider the \textit{context} that triggers a device to update, only that the device checked for an update. To further characterize update interaction, we look at event-related information to provide more context to the various conditions that cause IoT devices to update. All the packets captured from the Ren et al.\ study are labeled with various event-related information such as power events, app interaction, or idle events. Therefore, we analyze these crucial pieces of context to correlate events to update activity. For example, if an IoT device checks for an update when powered on, an adaptive firewall can use temporal data of an IoT device's network connectivity to provide more context to classify if an IoT device may be requesting and applying a software update.  

Finally, we extract and analyze all TLS handshake data from all the packet captures (independent of update keyword traffic) to assess the overall strength of the communication channels in use. 
Our methodology only allows us to perform extended analysis on unencrypted traffic; however, if IoT devices send all of their traffic over an encrypted medium, it is a reasonable assumption that the devices will also perform firmware updates over these encrypted connections. If the TLS implementation on the IoT device is outdated or insecure will undermine the overall security of the IoT device, including the software update system. Whether TLS is explicitly or implicitly chosen for a design, using TLS is a design choice for IoT update systems.

To interpret the set of cipher suites advertised between clients and servers, we converted the cipher suite's hexadecimal value to the IANA cipher suite name by leveraging a cipher suite information API~\cite{ciphersuiteinfo} which aggregates all IANA cipher suites along with IANA cipher suite security classifications. Cipher suites are then categorized into four buckets: insecure, weak, secure, and recommended. Insecure cipher suites have easily exploitable security flaws and thus should \textit{never} be used, while weak cipher suites may have proof-of-concept vulnerabilities that are more difficult to exploit in practice. The classes of secure and recommended cipher suites have no known vulnerabilities, and all recommended cipher suites are a subset of secure cipher suites. The only differentiating factor is that recommended cipher suites support Perfect Forward Secrecy (PFS).     \section{Results}\label{sec:Results}
In this section, we discuss our results in identifying update-related traffic. At the network level, software updates are difficult to detect if the update communications are taking place over an encrypted connection. TLS offloading may be an option in non-IoT contexts; however, attempting TLS offloading on IoT devices will require physically tampering with the device which may cause erratic behavior~\cite{ren_information_2019}. 

Our HTTP object extraction pipeline extracted HTTP objects from 5,766 of 38,356 packet captures, which is $~15\%$ of the packet captures in the dataset. In other words, $85\%$ of packet captures use some form of encryption, or a protocol other than HTTP. We extracted HTTP data for 35 out of 55 devices\footnote{Originally, Ren et al.\ had 81 devices with 26 common devices between regions, thus 55 unique devices.}, which is $63\%$ of devices. Originally, Ren et al.\ attempted to measure encryption adoption with slightly different results: no device had more than $75\%$ unencrypted traffic~\cite{ren_information_2019}. The key difference in our results is we focus on extractable HTTP objects, whereas Ren et al.\ attempted to guess if certain UDP traffic was encrypted or not by measuring byte entropy, which only concludes if certain packets are \textit{likely} encrypted~\cite{ren_information_2019}. 

In the following sections, we describe our results for identifying software update keywords, characterizing software updates based on device interaction, and our TLS results. These results are summarized as follows: 

\begin{itemize}
    \item \ref{sectionUpdateKeywordsResults}: Out of the 35 devices that did not encrypt all traffic, 9 ($25\%$) checked for available software updates transparently.
    
    \item \ref{sectionUpdateEventsResults}: Update-related traffic is correlated to power and idle events, but a small percentage of devices checked periodically (some as often as once per hour). 

    \item \ref{sectionDesignPatternResults}: Update endpoints (where software update files are hosted) for devices in our set exist primarily in 3rd party cloud service platforms, or on content delivery networks (CDNs), which makes DNS-based identification difficult.

    \item \ref{sectionCipherSuiteResults}: TLS is pervasively used in IoT communications, possibly including update-related traffic. Devices that only use TLS for communication could be vulnerable to key compromise if there are no additional protections in place~\cite{samuel_survivable_2010}. 
    \item \ref{sectionCipherSuiteResults}: The majority of our devices use secure TLS cipher suites which would not make them vulnerable to TLS downgrade attacks; however, there are devices that support vulnerable TLS cipher suites, which jeopardizes any update communications made through TLS.
\end{itemize}

\subsection{Update Keywords Results} \label{sectionUpdateKeywordsResults}
\begin{figure}[htp]
    \centering
    \subfloat[Update keyword occurrences by interaction event. A lighter color indicates a higher percentage of occurrences. In the original study~\cite{ren_information_2019} there were 9 different Alexa interaction events, and 4 android interaction events, which we chose to merge into a single group for readability.\label{eventKeywords}]{\includegraphics[width=0.47\textwidth]{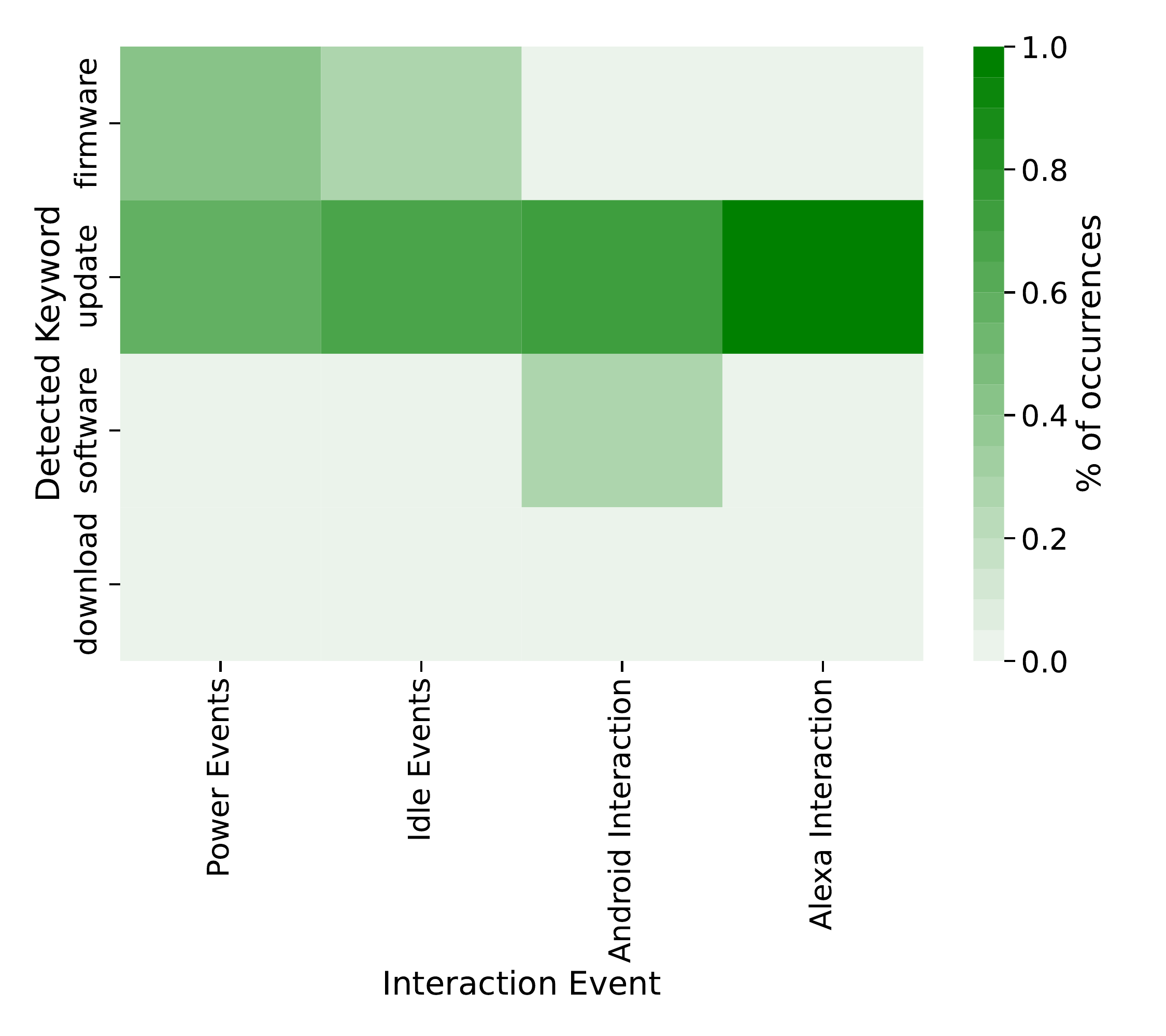}}\hfil \subfloat[Count of detected update keywords aggregated by device as a heatmap, where the number in the square corresponds to the number of keyword usage occurrences were found.\label{iotUpdateKeywordsByDevice}]{\includegraphics[width=0.47\textwidth]{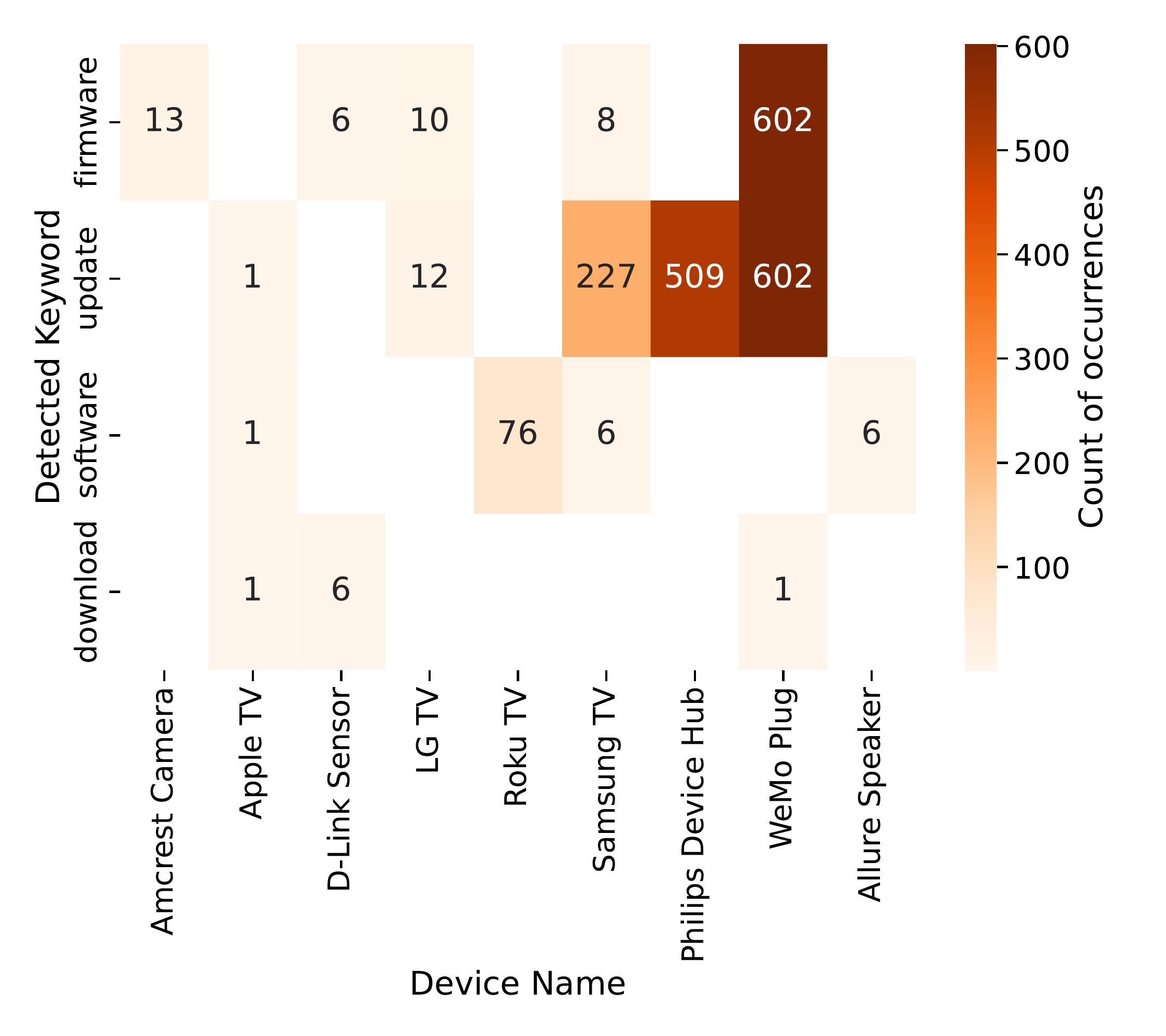}}

    \caption{Our results for update keywords by device and interaction event.}
\end{figure}
\begin{figure}
    \centering
    \includegraphics[width=\textwidth,height=3in]{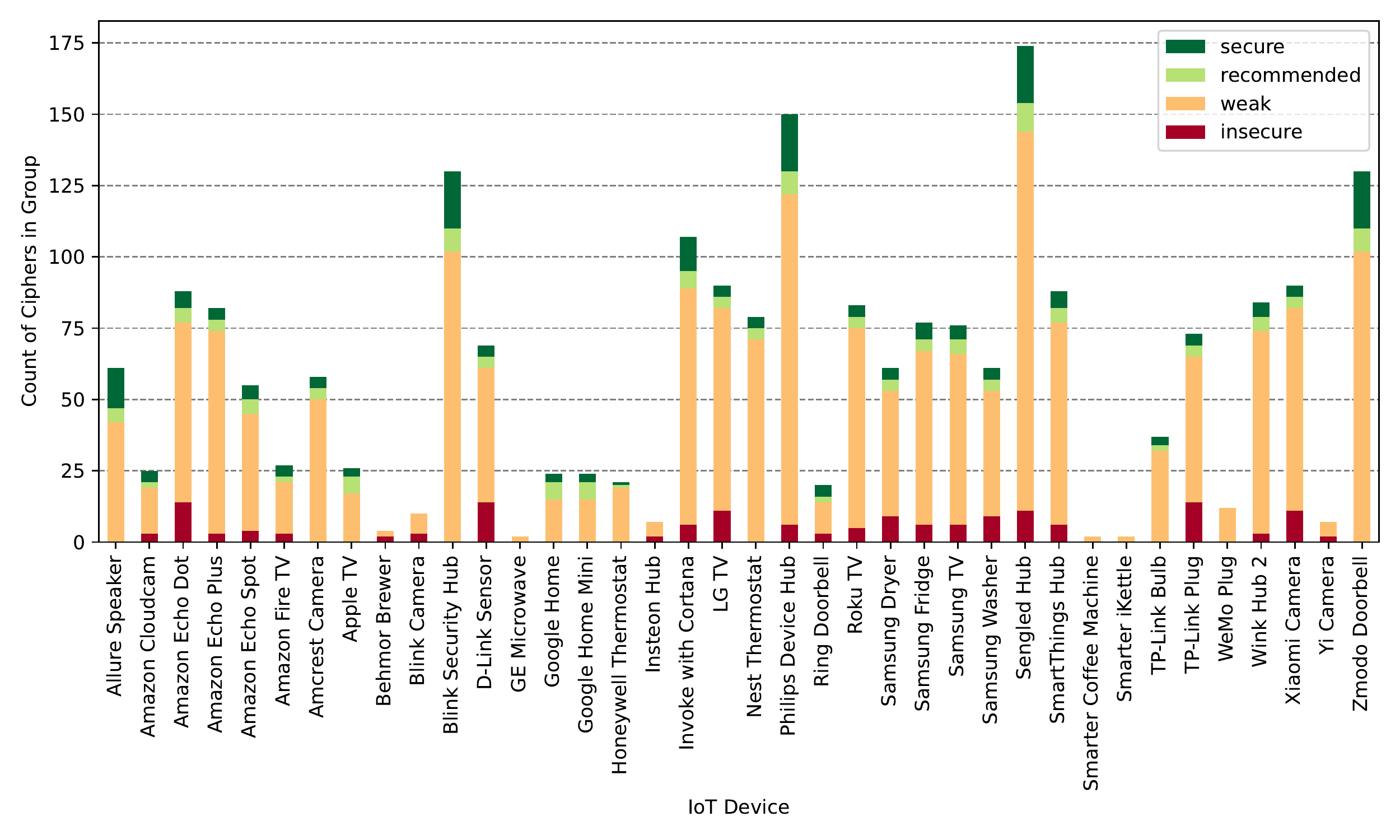}
    \caption{Count of TLS cipher usage on a per-device basis. Each bar is represented by insecure, weak, secure, and recommended cipher suites.}
    \label{tlsCipherByDevice}
\end{figure}

We successfully extracted several HTTP interactions between IoT devices and web services related to software updates. Our most prominent keyword is \textit{update} with 1,351 occurrences among extracted HTTP objects, \textit{firmware} with 639 occurrences, \textit{software} with 89, and \textit{download} appearing only 8 times.

The specific devices and the corresponding keywords they matched are shown in Figure~\ref{iotUpdateKeywordsByDevice}. The heatmap shows the number of occurrences of the keywords in the rows for the devices in the columns, where a darker blue indicates more occurrences. We observed that certain devices exchange update-related information much more often than others, such as the Wemo plug and Phillips hub. 

The Wemo plug device had the most occurrences of keywords, which means the Wemo plug was polling the most frequently for updates; however, this does not imply there may be a software update in progress. For example, the Wemo Plug exchanges firmware information in nearly every request which contributes to the high amount of keyword detection; however, we did not find any proof that the Wemo plug performed an update during the capture period. There is an update web service offered by the Wemo plug, which we discuss in detail in Section~\ref{sec:wemo}. By contrast, the Apple TV only has a single occurrence of exchanging update-related keywords, and we found that the Apple TV downloaded system firmware over HTTP, which would imply that the Apple TV installed the aforementioned firmware, which we discuss in Section~\ref{sec:appletv}. This contrast shows that our heuristic does not guarantee a device is performing an update, but it is enough to detect traffic that \textit{might} be update related.

Aside from being able to detect firmware downloads in real-time, an unexpected result from our heuristic was it picks up current updates and firmware versions in 7 of the 9 devices. This is because these 7 devices report their firmware version as an HTTP request, or as part of a service discovery response. This is valuable information for both defensive and offensive applications. A potential application for this in defensive security is an active firewall appliance that can scan IoT devices and fetch firmware versions from them, if a CVE is released for that particular firmware the firewall can automatically quarantine the affected devices. This assumes that the firmware version is accurately reported, which may not be the case for malicious devices. For offensive security applications, an attacker could perform reconnaissance by identifying vulnerable firmware versions of devices that actively advertise these versions.

\subsection{Update Events Results} \label{sectionUpdateEventsResults}

Our results for event-related update activity are shown in Figure~\ref{eventKeywords}. The heatmap shows the number of update keyword occurrences in the rows for the interaction event in the columns, where a lighter color indicates more occurrences. Due to the granularity of the experiments from Ren et al., Android-related events (e.g., taking a photo, controlling a device from an app, etc.) and Alexa interactions (e.g., invoking Alexa, changing color, etc.) were merged into two respective categories. Aside from these events, all 9 of the IoT devices in Figure~\ref{iotUpdateKeywordsByDevice} exchange update-related keywords on power events, and even more on idle events. Examples of update traffic events include devices reporting their \textit{firmware} version to an update service, then receiving an \textit{update} response in return.

When idle, we found some IoT devices that exchange update-related traffic between one another. This is out of the ordinary, as independent IoT devices should not be issuing or exchanging update commands to one another when idle -- these communications should only occur between the device and the vendor's update platform. We investigated these inter-device occurrences and found that as part of service discovery protocols (e.g., SSDP, UPnP) there is an exchange of firmware information. Certain devices even advertise endpoints for invoking update behavior manually which is ripe for exploit by bad actors or rogue IoT devices. Section~\ref{sec:wemo} for more information regarding these endpoints.

Other than power and idle events, Alexa interaction events contribute the most to our heatmap. Alexa devices do not exchange detectable update-related traffic; however, the Philips hub exchanged update-related information when being controlled by Alexa. Additionally, the Roku TV, Samsung TV, and Wemo plug exchanged update-related data when controlled remotely by Android interaction events. We believe there is no correlation between these interactions and update traffic: these devices exchange the same information when not being controlled by Alexa or Android.

\subsection{Observed Update Design Patterns} \label{sectionDesignPatternResults}

We analyzed the extracted HTTP interactions flagged as being update-related to attempt characterizing common designs or behaviors between device vendors. Unfortunately, no common architecture or strategy was used between the 9 devices we identified. The heterogeneity of the designs and schemas involved provide great motivation for standardized update system designs, such as RFC 9019 and RFC 9124~\cite{rfc9019,rfc9124}. While there is no common schema among different device vendors, we noticed some common patterns among certain device manufacturers. 

\designpattern{pattern:nosec}{\textbf{No Security}}: The D-Link movement sensor, Amcrest camera, and Wemo fetch firmware update metadata from a web service that returns a complete URL for downloading the firmware image. What is concerning about this is there is no tamper-protection in place for any of these devices. To make matters worse, both of these devices fetch data from public S3 bucket endpoints over HTTP.
We examined firmware images served through these endpoints and found no forms of tamper-protection such as checksums, digital signatures, etc. built into the firmware. 

\designpattern{pattern:outofband}{\textbf{Out-of-band Security}}: While insecure device update schemes are certainly concerning, there are update techniques that allow authentication and integrity verification even over HTTP. The Apple TV exchanged all update-related traffic over HTTP, including web service interactions for downloading the firmware and related metadata. What sets the Apple TV apart is it exchanges digital signatures and certificates over HTTP to validate the responses. Apple's design provides a happy medium of ensuring the integrity (assuming the signatures and certificates are validated) of the update through cryptographic means while giving us insight into specific details that can be leveraged by a network appliance, such as specific firmware and information assuming that the network appliance can parse the XML schema Apple uses.

\designpattern{pattern:fulltls}{\textbf{Full TLS}}: The remaining devices encrypted all cloud-destined communications using TLS. It is reasonable to expect that, if implemented, a software update mechanism would also use one of the available TLS channels. While communication encryption is advantageous for security and privacy, we believe transparency in software update implementations (perhaps implemented with an out-of-band scheme as described above) can be beneficial for providing transparency and security, as we described in Section~\ref{sec:intro}. Additionally, we note that exclusive reliance on TLS for software updates is known to be insufficient in protecting against many update-specific attacks~\cite{samuel_survivable_2010}.

\subsection{Cipher Suite Results} \label{sectionCipherSuiteResults}

We see a larger amount of devices with extractable TLS cipher suites, which is expected as many IoT devices use TLS as a means of interacting with the web services they depend on. In Figure~\ref{tlsCipherByDevice} we observe there were a total of 16 insecure cipher suites used between IoT devices. All 16 cipher suites have significant vulnerabilities that when combined with a downgrade attack could allow an attacker to perform a machine-in-the-middle (MITM) attack; however, among the 24 devices that advertise insecure cipher suites, we estimate 4 of them would be vulnerable to a downgrade attack. This is because the \textit{secure} and \textit{recommended} cipher suites would take precedence over the weak and insecure cipher suites, and the cipher suites contained in secure and recommended classes contain measures to prevent downgrade attacks. 

We have only discussed the TLS cipher suites in the context of IoT devices. To see these results in perspective to other applications that require secure communication, we searched for a dataset of TLS cipher suite support in web browsers. While we did not find a comprehensive dataset that summarized recent browsers, we did find a service that provides us with what our browser supports~\cite{ssllabs}. Using this service, we found modern browsers (Firefox 94, Chromium 96) support far fewer cipher suites with none of them being insecure -- although roughly half of the cipher suites supported were deemed to be ``weak''. This can be used to offset the large amount of IoT devices that also offer large amounts of ``weak'' cipher suites, as these may only be present for backward compatibility. In this context, the weak cipher suites used by IoT devices do not strictly increase the attack surface as compared to modern web browsers; however, insecure cipher suites when not using TLS 1.3 do increase the attack surface.

\subsection{Limitations} \label{sectionAnalysisLimitation}

We found 11 devices that did not have extractable HTTP data or extractable TLS data. By manually inspecting packet captures we found several devices that stream data over UDP, which is a consistent finding with the Ren et al.\ study~\cite{ren_information_2019}. The data was not meaningful, as it was either encoded using some vendor-specific encoding or a stream of application-specific data (e.g., a video stream) that can not easily be deciphered. While these edge cases are technically possible to extract, it is challenging to do so at scale given the wide breadth of devices and a large amount of packet captures.

A limitation of our study is TLS encrypted traffic, which is consistent with other large scale IoT analysis papers~\cite{prakash_software_2022,paracha2021iotls}. A potential workaround for TLS-encrypted edge cases is an alternative heuristic: for example, another approach that is agnostic to the protocol in use is to look at response sizes. If a device exchanges a large amount of data in a short burst, assuming that this burst of traffic is abnormal for the device based on regulapproach is not ideal as there is no way to verify if traffic is update-related -- this only identifies large bursts of abnormal traffic. Furthermore, even if we could deduce that encrypted traffic is a device update, there is no meaningful extractable information from an encrypted payload such as firmware version which is crucial to our motivation for detecting IoT software updates.

Another potential heuristic is to analyze traffic patterns temporally. O'Connor et al.\ developed a simple yet effective methodology for classifying various IoT subsystems without any form of decrypting or inspecting packet payloads, instead opting to analyze traffic frequency and size over a long period of time~\cite{oconnor_blinded_2019}. This temporal approach proved effective for identifying IoT device telemetry, and in an active measurement context, O'Connor et al.\ were able to derive various attacks based on a temporal analysis of IoT device traffic. While this approach is novel, it is not ideal for a large-scale passive analysis of traffic.

Regarding the keyword-based analysis, our heuristic which associates terms such as ``firmware'' and ``software'' to update-related events can produce false positives. For example, some devices report a current firmware version to a web service contained as an HTTP payload. While this is not an update request, our pipeline will flag it as such and require manual removal. Future work will investigate the use of additional heuristics to improve the accuracy of identification of updates without requiring manual verification. Adding checks for outbound traffic, inbound traffic, and schema verification would greatly assist in avoiding false positives.     \section{Case Studies}

In this section, we discuss our findings by analyzing select update practices and firmware files that we extracted through our methodology. First, we look at the firmware update interactions from the D-Link Camera, which we use to illustrate harmful practices that undermine the device's security. We then contrast this approach with the firmware update interactions we observed against the Apple TV, which combines several distinct tamper-resistant mechanisms with update transparency. Finally, we conclude our case studies with a vulnerable WeMo update service, that allows for unsigned code to be uploaded from an arbitrary source.

\subsection{D-Link Camera Firmware}~\label{sec:dlink}
The D-Link camera is an example of the~\ref{pattern:nosec} pattern, as it exchanged firmware update information through HTTP. Based on the identified traffic, we extracted a firmware update endpoint and also a firmware image. The firmware update endpoint is a web service that accepts a device model and returns an XML response containing firmware metadata information along with a URL to the latest firmware download. We were able to download the latest firmware image as it is being hosted by a static file store which does not require any prior authorization. The firmware update endpoint does not return any checksum or signature to validate that the firmware image was not tampered with. Using the \texttt{binwalk} utility\footnote{\href{https://github.com/ReFirmLabs/binwalk}{https://github.com/ReFirmLabs/binwalk}} we analyzed the firmware image and found the following:

\begin{enumerate}
    \item A $\mu$Image header, indicating that the OS is Linux built for a MIPS CPU. This is likely a boot loader for the next item
    \item LZMA compressed data, likely the kernel image to be executed by (1)
    \item A SquashFS filesystem, which is the root filesystem
\end{enumerate}

The image header indicates that the OS is a Linux Kernel from roughly 2014 (6 years old at the time of writing). Looking at the kernel image (2) we extracted the image version, which is Kernel version \texttt{2.6.31} released in 2009~\cite{torvalds_2009}. While we did not find any notable CVEs for this particular version (\textit{2.6.31}) of the  kernel~\cite{cve_kernel_231}, we did find CVEs for the parent minor version (\textit{2.6}) which allow for arbitrary code execution through multiple buffer overflows~\cite{cve_2008_4395}. 
It is likely after 2014 the device reached the end of its ``service life'', thus D-Link stopped updating it. This is unfortunately a fairly common occurrence amongst IoT devices~\cite{rahman_understanding_2018}.

 Theoretically speaking, the D-Link camera is vulnerable to MITM attacks as shown in Figure~\ref{insecureComms}: the communication between (1) the update service and (2) the image repository is unauthenticated and does not have any integrity protection. For (1), an on-path attacker can intercept traffic between the IoT device and the vendor's cloud. In this case, the message responded by the vendor's cloud contains the full URL to the firmware image being hosted on an S3 bucket (also on HTTP). A second MITM attack (2) could occur if an attacker intercepts HTTP traffic between the IoT device and the S3 bucket. With this in mind, it is highly likely an attacker can leverage (1) to give the D-Link camera the URL of a different S3 bucket hosted on the ``malicious cloud instance'' which would then serve the modified firmware. An attacker could build and distribute modified firmware trivially, as the original firmware file is not signed digitally or otherwise clearly authenticated.

\subsection{Apple TV Firmware}~\label{sec:appletv}

\begin{figure}[htp]
    \centering
    \subfloat[AppleTV update process between Apple's content delivery network and update authorization server. There are two distinct stages to the update process: first downloading the update bundle via the Apple Updates CDN, then validating and authorizing the update through remote attestation.\label{appleUpdateFlow}]{\includegraphics[width=0.45\textwidth,height=2.26in]{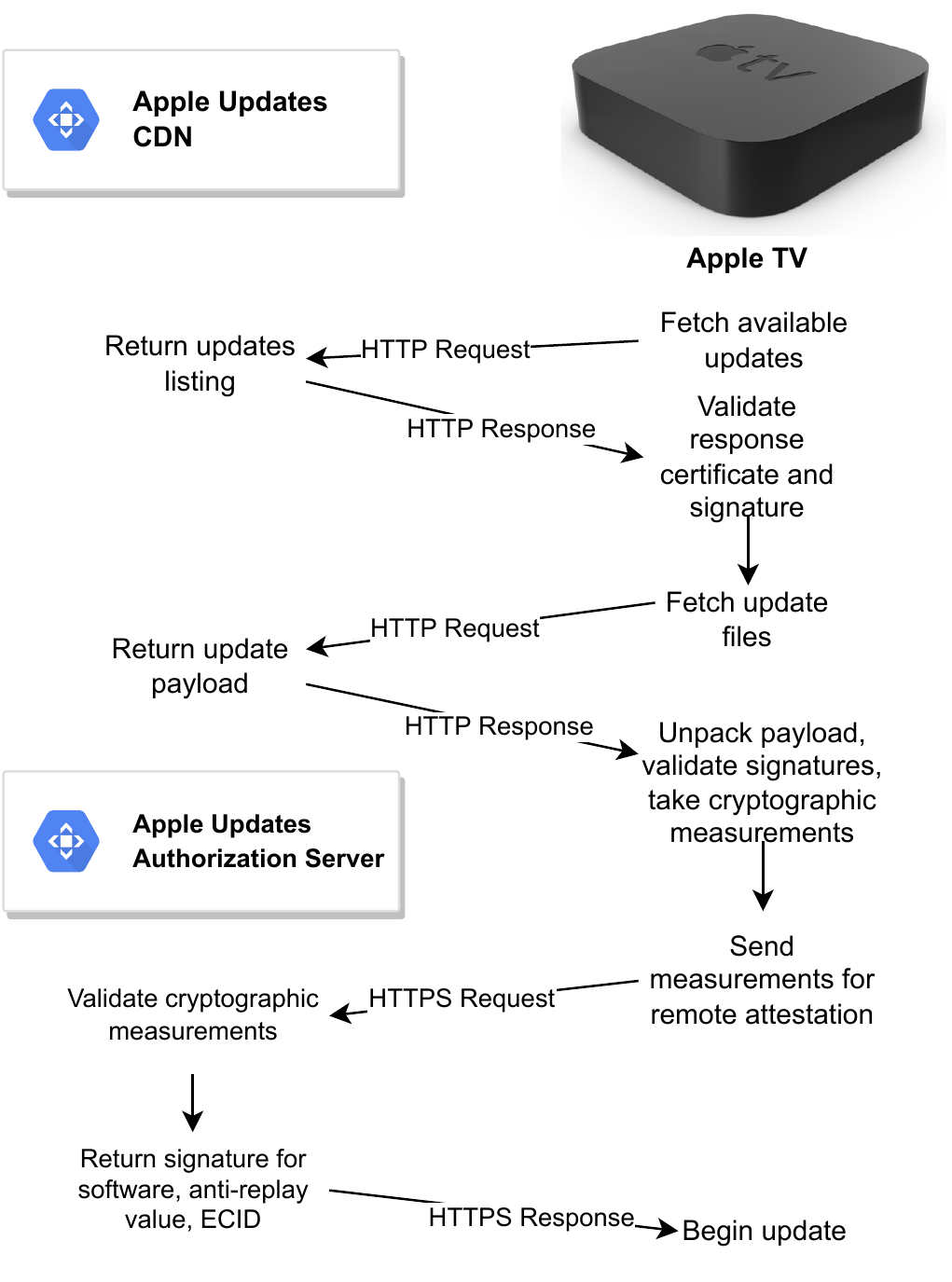}}\hfil
    \subfloat[An example MITM attack scenario that the D-Link camera is vulnerable to. The attacker would appear as an authentic source that provides a malicious payload, such as a download link to a modified firmware version being hosted by the attacker. \label{insecureComms}]{\includegraphics[width=0.45\textwidth]{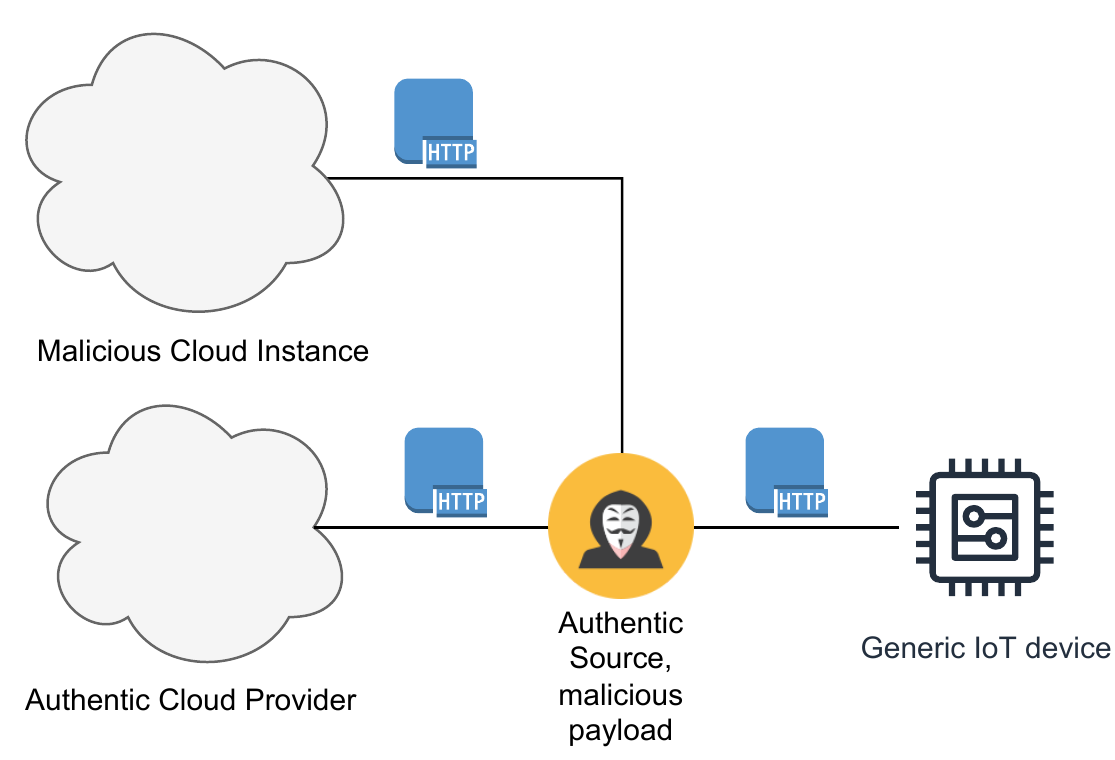}}
    
    \caption{Our results for update keywords by device and interaction event.}
\end{figure}

In a contrast to the D-Link camera, the Apple TV's update behavior combines security and transparency, making it an instance of the~\ref{pattern:outofband} pattern. The complete update flow of the Apple TV is shown in Figure~\ref{appleUpdateFlow}. Similar to the D-Link Camera update metadata is exchanged over HTTP; however, there are several additional measures to harden communications against attackers. 

The Apple TV first connects to a central update repository over HTTP. Although the connection for update metadata happens over HTTP, we found the API response contains a certificate and signature field, which is used to validate the responses integrity~\cite{apple_updatesecurity}. We found the certificate was issued by the  ``Apple iPhone Certification Authority'', with a common name of ``Asset Manifest Signing''. This suggests that the certificate is purpose-made specifically for signing these update manifest responses. Unfortunately, the certificate expired in 2018, and the API response indicated updates from as recently as 2020.

Downloading the update files also takes place over HTTP. To protect against tampering there is an additional field containing a validation measurement for the update file. If the update file is downloaded and does not match the measurement, the update is invalid and rejected. This behavior is consistent with Apple's platform security documentation which details the measures taken to secure device updates~\cite{apple_updatesecurity}.

Using the Apple Repository response, we reconstructed the firmware download URL and acquired the firmware image for the Apple TV. When unpacked, the firmware contains a file tree for distributing software updates. Without having the source code to the software responsible for performing updates on Apple devices, we are unable to determine how exactly the update is performed; however, combining an analysis of the directory tree with prior reverse engineering efforts~\cite{iphonewiki_ota} along with Apple's platform security documentation~\cite{apple_updatesecurity} gives us good insight into how the update is performed past this point.

After the AppleTV validates the update payload, the AppleTV must perform remote attestation with the Apple Updates Authorization server to fetch keys that are required to perform the update. According to our packet captures, this communication takes place over HTTPS (as pictured in Figure~\ref{appleUpdateFlow}), so we do not have concrete knowledge of what exactly is being exchanged. According to Apple's platform security documentation, cryptographic measurements of the bootloader (iBoot), kernel, operating system image, and exclusive chip ID (ECID) are sent to the update authorization server~\cite{apple_updatesecurity}. The server validates all the measurements sent by the device, and if they are valid, the update server returns the signature for the software, an anti-replay value, and the device's ECID~\cite{apple_updatesecurity}.

\subsection{WeMo Update Service}~\label{sec:wemo}

The Belkin WeMo plug largely communicates using Simple Service Discovery Protocol (SSDP), which is a protocol used to advertise services and consume them in a standardized way~\cite{cai-ssdp-v1-03}. SSDP uses HTTP as its underlying communication protocol, therefore all SSDP activity was captured by our passive analysis. We observed amongst the various device management services listed is one for firmware updates. The firmware update service advertised various methods for firmware management, one of particular interest is the ``UpdateFirmware'' method, which accepts various parameters describing the new firmware -- one such parameter allows for an unsigned image to be uploaded, which has been historically shown to be exploitable~\cite{cvewemo,buentello_2013}. An attacker could have a local or remote firmware repository and upload a modified firmware image to the device. Due to the lack of authentication and authorization on this SSDP endpoint, this is an instance of the~\ref{pattern:nosec} pattern.

We cannot test the viability of uploading arbitrary firmware to the WeMo update service as we are passively analyzing packet captures; however, previous efforts aimed at exploiting this update endpoint have proven to be successful, leading to arbitrary firmware uploads to the WeMo device~\cite{buentello_2013}. An attacker could have a local (or remote) firmware repository, and upload a modified firmware image to the WeMo device. The only difference between the exploit used in the D-Link camera and the WeMo plug is the attacker has the ability to trigger device update behavior by interacting with an endpoint, whereas the D-Link camera has no such functionality.
     \section{Related Work}

To our knowledge, this is the first work attempting to analyze and characterize how consumer IoT devices perform software updates at the network level. There have been recent works focusing on the different network-level analysis of IoT devices: Prakash et al.\ analyze the update practices of IoT vendors by tracking software versions listed in the user-agent header included in HTTP requests made by IoT devices~\cite{prakash_software_2022}. The conclusions found by Parakash et al.\ do not characterize and analyze how IoT update systems work, rather, they conclude that IoT device vendors are slow to update their devices when new vulnerabilities are found.

We identified pervasive use of TLS, which precludes the identification of update-related traffic without additional data analysis. Related work here includes Alrawi et al., who provide an excellent SoK of the overall security of home IoT devices by systematizing the current state (as of 2019) of IoT vulnerability literature and then evaluating 45 devices, a subset of the security evaluation involves looking at various encryption qualities that would make the device vulnerable~\cite{tlsSOK}. More recently in 2021, Paracha et al.\ performed a deep dive into IoT TLS usage patterns which ultimately found 11/32 IoT devices are vulnerable to interception attacks~\cite{paracha2021iotls}. If IoT devices are relying on TLS to secure communications to backend APIs and endpoints for software updates, any vulnerabilities in the TLS transport layer will undermine the overall soundness of how these devices perform updates.

An encouraging finding is the high amounts of TLS usage among devices; however, there is a caveat to this high TLS usage: it is only one line of defense. If a private key is compromised, this could jeopardize the integrity of update-related services if there are no additional lines of defense. Samuel et al. present a novel design for an updated system that allows for key compromise in update systems~\cite{samuel_survivable_2010}.

Due to the previously discussed challenges, there are several opportunities to explore and innovate IoT software update designs. Related work in this space consists of proposed designs for IoT update systems relating to firmware updates and library management. Zandberg et al.\ present a prototype for a firmware update system on IoT devices by leveraging various open-source libraries and standards~\cite{zandberg_secure_2019}. Zandberg et al.\ leverage SUIT, a new IETF standard that provides \textit{encrypted} firmware update files with encryption keys provided by hybrid public-key encryption~\cite{ietf-suit-firmware-encryption-02}. The SUIT standard appears as if it may not work on resource-constrained IoT devices, but Zandberg et al.\ have their reference implementation built on IoT devices with less than 32 KB of RAM and 128 KB of storage~\cite{zandberg_secure_2019}.     \section{Conclusion}

Using a passive measurements approach and a dataset from one of the largest IoT information exposure studies to date~\cite{ren_information_2019} we identified and characterized several design patterns used by IoT devices to perform updates. There is no common schema or design pattern behind various update systems, which provides additional motivation for standardizing IoT software updates~\cite{rfc9019}. Additionally, we characterized  events related to when an IoT device may update, which is useful for building data-driven models for real-time update identification. In our analysis of update systems, we found vulnerable devices that provide no mechanisms for securing firmware updates. We observed that many devices use encrypted connections to secure communications: 60\% of devices support insecure TLS cipher suites, while 10\% of devices are vulnerable to downgrade attacks.

In the future, more comprehensive studies can follow by performing active measurements during software updates. This can reveal more IoT update endpoints, allow us to develop more accurate heuristics for identifying when a device is updating, and therefore gain a better understanding of these walled gardens. 

\section*{Acknowledgements}
    This research is supported by the Natural Sciences and Engineering Research 
    Council of Canada (NSERC) through a Discovery Grant.
    
\bibliographystyle{splncs04}
    
    \end{document}